# CHARACTERIZATION OF THE MICROSTRUCTURE OF ZIRCONOLITE-BASED GLASS-CERAMICS

P. Loiseau [1], I. Touet [2], D. Caurant [1], Y. Dextre [2], and C. Fillet [3]

[1] ENSCP-LCAES, 11 rue P. et M. Curie, F-75231 Paris, Cedex 05
[2] CEA-CEREM, CEA Grenoble, 17 rue des Martyrs, F-38054 Grenoble, Cedex 9
[3] CEA-DCC, CEA-Valrhô, BP171, F-30207 Bagnols sur Ceze, Cedex

**ABSTRACT**

December 1991 legislation in France has spurred research on enhanced separation and conditioning or transmutation of long-lived radionuclides from high level radioactive wastes (HLW). In this field, we have studied zirconolite-based glass-ceramics in which the crystalline phase (zirconolite: $CaZrTi_2O_7$) aimed to preferentially incorporate minor actinides is embedded in a glassy calcium aluminosilicate matrix. At the laboratory scale, the crystallization of the parent glass is carried out thanks to a two-step thermal treatment: a nucleation stage followed by a growth stage. This paper presents the evolution of the crystallization, followed by scanning electron microscopy (SEM) and X-ray diffraction (XRD), with the temperature of the crystal growth thermal treatment, in the range 950° – 1350°C.

**INTRODUCTION**

Research about new conditioning matrices for minor actinides (Np, Am, Cm), which would result from an enhanced separation of long-lived radionuclides from HLW, are currently investigated. The objective of this study is to develop zirconolite-based glass-ceramics that would preferentially immobilize minor actinides in the zirconolite crystalline phase. Thus a double containment is reached thanks to a homogeneous dispersion of the crystals in the glassy matrix. Zirconolite is a crystalline phase well known for its excellent containment capacity and long-term behavior (high chemical durability and self-radiation resistance) (1). The results presented in this paper deal with microstructural analysis of zirconolite-based glass-ceramics prepared by heat treatment of a parent glass. The influence of crystal growth thermal treatment temperature $T_c$ (ranging between 950° and 1350°C) on the microstructure of the materials was studied. The different crystalline phases formed after devitrification were characterized by their composition, size, shape, crystalline structure and their distribution in the residual glassy matrix.

**SAMPLE PREPARATION AND EXPERIMENTAL METHODS**

The glass composition studied is given in Table 1. It is based on the formulation established by the CEA-Marcoule (2, 3). All the samples were prepared using the following operating mode :
- melting at 1550°C in a platinum melting pot (10h)
- pouring in water and grinding for glass homogenization before remelting at 1550° C (4h)
- casting and quenching at room temperature on a metallic plate
- thermal treatment at 810°C (slightly higher than the glass transformation temperature $T_g \approx$ 775°C) for 2h: nucleation stage
- thermal treatment at $T_c$ (950°-1350°C) for 2h: growth stage

Table 1. Composition of the studied parent glass (weight %)

| Oxides | $SiO_2$ | $Al_2O_3$ | $ZrO_2$ | CaO | $TiO_2$ | $Na_2O$ |
|---|---|---|---|---|---|---|
| % | 43.16 | 12.71 | 9.00 | 20.88 | 13.25 | 1.00 |

In order to study the intrinsic devitrification behavior of the parent glass, neither actinides nor simulants (lanthanides) were added to the composition. Results concerning the introduction of neodymium as actinide simulant are presented in other papers (4-6). Moreover these previous works have shown the occurrence of an opaque "crust" (a crystallized layer never exceeding 1 mm thickness) on the surface of the samples, while zirconolite crystals are homogeneously dispersed in the residual glass of the bulk for $T_c$ = 1000°-1200°C. Zirconolite probably forms in the bulk by homogeneous nucleation. The "crust" mainly consists of titanite ($CaTiSiO_5$) and anorthite ($CaAl_2Si_2O_8$) crystals that have nucleated heterogeneously on glass surface.

SEM observations and elementary analysis were performed on cut and polished glass-ceramic sections using a JEOL 840A microscope coupled to an EDS PGT/IMIX analysis system. Powder XRD patterns of the bulk of glass-ceramics were recorded in θ/2θ mode using a Siemens D5000 instrument with CoKα wavelength (λ = 1.78897Å).

**RESULTS AND DISCUSSION**

SEM images and XRD patterns corresponding to the bulk of heat-treated samples are shown in Figures 1 and 2 respectively.

After nucleation (810°C 2h), the glass remains transparent; SEM and XRD reveal no microcrystallization in the bulk and on the surface of samples. This indicates that during this step, crystal growth rate is not significant.

For $T_c$ ranging from 950° to 1200°C a high density of star-shaped crystals (5-10 μm diameter) have nucleated and grown in the bulk. These crystals are homogeneously distributed and were identified by XRD:

- For $T_c$ = 950°C, dendritic crystals have grown. XRD patterns indicate that they consist of a mixture of zirconolite and fluorite type phases. However only one kind of crystal morphology is observed by SEM. As there is no more fluorite type phase for $T_c$ higher than 950°C, it indicates that fluorite transforms into zirconolite at 950°C. These two phases probably differ only by cationic ordering (7).

- For $T_c$ ranging from 1000° to 1200°C, zirconolite crystal morphology changes progressively. The dendritic microstructure which consists of main branches (corresponding to preferred growth directions) and secondary branches clearly becomes simplified for $T_c$ greater than or equal to 1100°C. Disappearance of these secondary branches can be associated with a progressive decrease of the residual supercooled liquid viscosity and consequently with a lowering of concentration gradients near the surface of the main branches (non-congruent crystallization). Interface stability is thus expected to increase with $T_c$. At the same time, thickness of the main branches increases and a more open particle microstructure is observed. Moreover, XRD patterns evolution indicates that as $T_c$ increases, zirconolite lattice parameters progressively change (a splitting is observed for several XRD lines notably at 2θ = 35.5°, 42° and 59.5°). This evolution can be attributed to a progressive ordering of cations in the (Ca, Zr) planes of zirconolite (5).

From $T_c$ = 1250°C, a new crystalline phase identified as m-$ZrO_2$ (baddeleyite) forms at the expense of zirconolite which progressively disappears with increasing $T_c$. These crystals appear as white elongated globular particles on backscattered SEM images. The strong contrast evolution between residual glass, zirconolite and baddeleyite is due to an increase of zirconium content in the corresponding phases. EDS analysis indicates that the baddeleyite crystals contain significant $TiO_2$ amount. A low CaO level is also detected. These results are in accordance with the existence of relatively wide solid solution domains between $ZrO_2$ on the one hand, and $TiO_2$ and CaO on the other hand (8,9). For $T_c$ = 1350°C, almost all zirconolite crystals have disappeared. They remain only as very thin needles decorating $ZrO_2$ particles. This shows that the dissolution limit of zirconolite is slightly higher than 1350°C. The fact that relatively large $ZrO_2$ crystals remain at this temperature indicates that this phase dictates the liquidus temperature of the system. It ban be noticed that the occurrence of baddeleyite at high temperature at the expense of zirconolite is in accordance with the phase diagram evolution published by Xu et al. (10) for the pseudo-binary $ZrO_2$ - ($TiO_2$ + $CaTiO_3$) system.

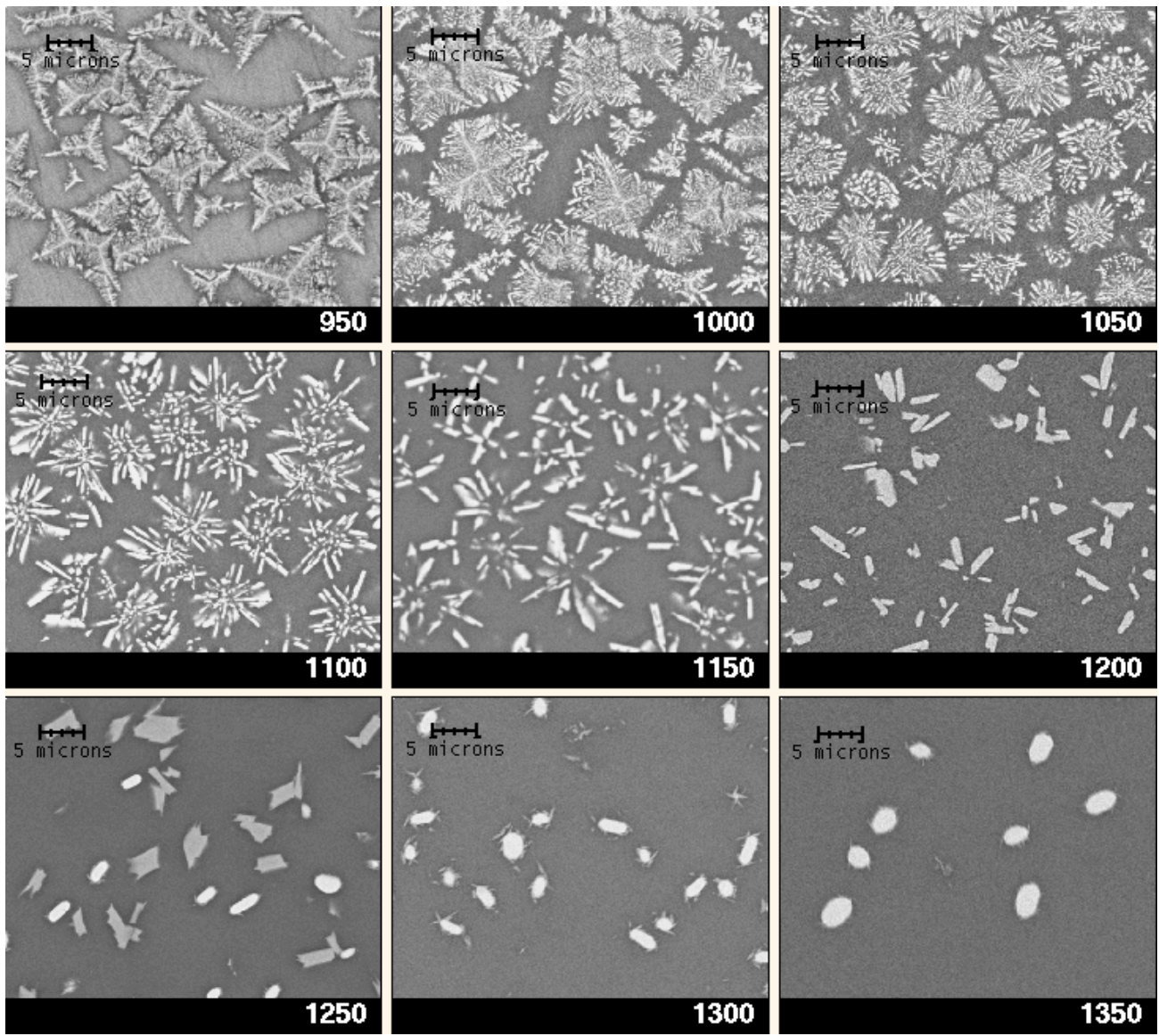

**Figure 1. Backscattered SEM images of the bulk of the glass-ceramics prepared at the indicated crystal growth temperatures $T_c$ (x 3000)**

The crystallized layer forming a "crust" on sample surface (not shown in Figure 1) is observed for all the samples prepared for $T_c$ lower than 1300°C. Its thickness progressively grows until 1150°-1200°C and then strongly drops as to disappear from $T_c$ = 1300°C.

## CONCLUSION

The effect of crystal growth temperature $T_c$ on the microstructure and the structure of the crystalline phases formed in zirconolite-based glass-ceramics designed as durable waste forms for minor actinides immobilization was investigated. For $T_c$ ranging from 1000° to 1200°C, zirconolite is the only crystalline phase growing in the bulk. The cationic order in zirconolite crystals increases with $T_c$ and the dendritic microstructure observed for low $T_c$ progressively disappears. For $T_c$ higher than 1200°C, $ZrO_2$ crystals grow in the bulk at the expense of zirconolite and dictate the liquidus temperature of the system. Thus there is a 200°C safety margin for the preparation of glass-ceramics with zirconolite as the only crystalline phase in the bulk.

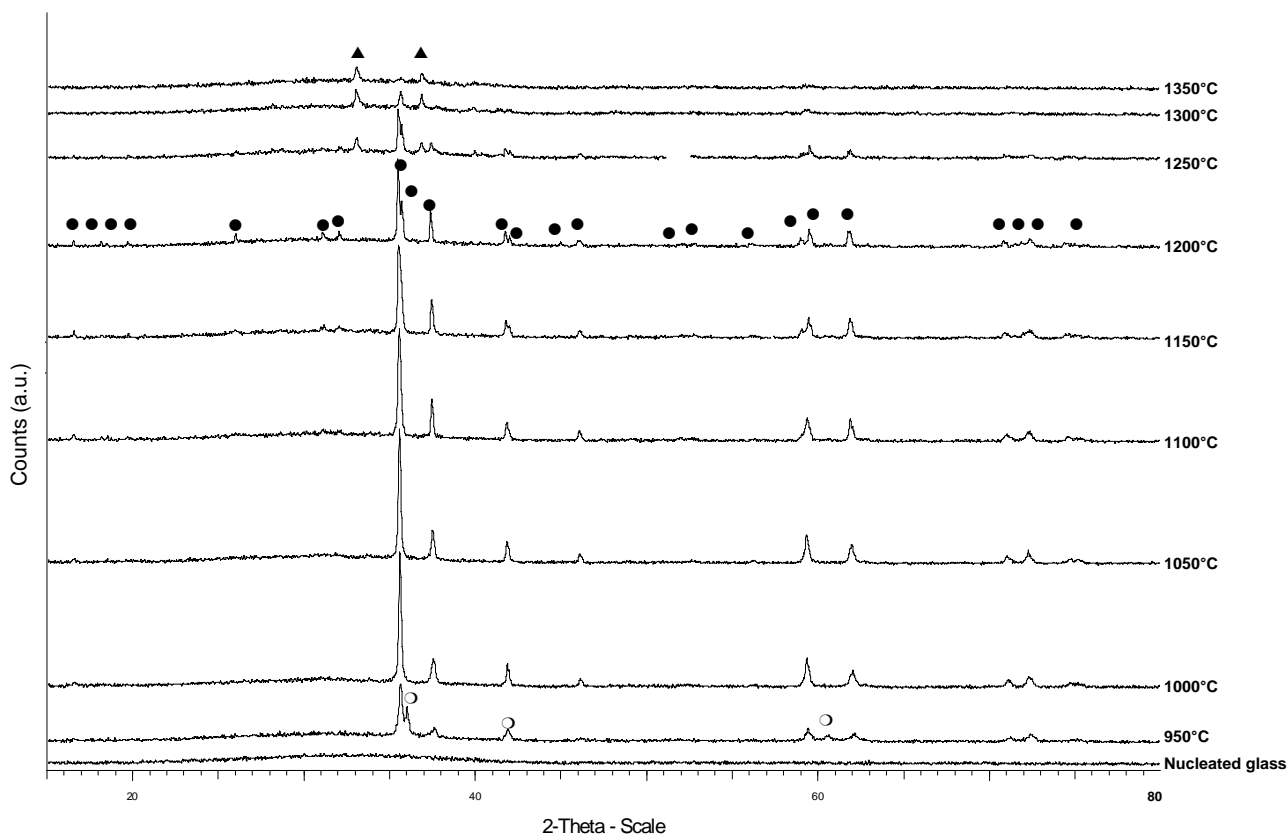

**Figure 2. XRD patterns of the nucleated glass and of the bulk of the glass-ceramics prepared at $T_c$ ranging from 950° to 1350°C (○ fluorite type phase, ● zirconolite, ▲ baddeleyite)**

**REFERENCES**

(1)  P. E. Fielding, T. J. White, J. Mater. Res. **2**(3), 387 (1987)
(2)  C. Fillet, J. Marillet, J.L. Dussossoy, F. Pacaud, N. Jacquet-Francillon, J. Phalippou, Environmental Issues and wastes Management Technologies in the Ceramic and nuclear Industries III, **87**, 531 (1997)
(3)  T. Advocat, C. Fillet, J. Marillet, G. Leturcq, J.M. Boubals, A Bonnetier, MRS Symp. Proceedings, SBNWM XXI, 55 (1998)
(4)  P. Loiseau, D. Caurant, N. Baffier, C. Fillet, Atalante 2000 (this issue)
(5)  P. Loiseau, D. Caurant, N. Baffier, L Mazerolles, C. Fillet, MRS Symp. Proceedings, SBNWM XXIV (2000) (to be published)
(6)  P. Loiseau, D. Caurant, N. Baffier, C. Fillet, MRS Symp. Proceedings, SBNWM XXIV (2000) (to be published)
(7)  E. R. Vance, C. J. Ball, M. G. Blackford, D. J. Cassidy, K. L. Smith, J. Nucl. Mater. **175**, 58 (1990)
(8)  F. H. Brown, P. Duwez, J. Am. Ceram. Soc. **37**(3), 132 (1954)
(9)  P. Duwez, F. Odell, F. H. Brown, J. Am. Ceram. Soc. **35**(5), 109 (1952)
(10) H. Xu, Y. Wang, J. Nucl. Mater. **279**, 100 (2000)